\documentclass[journal]{IEEEtran}
\usepackage{amssymb,amsmath,algorithm,algorithmicx,algpseudocode,epsfig,microtype,diagbox,multirow,color}

\newcommand\Algcase[1]{%
\vspace*{-.7\baselineskip}\Statex\hspace*{\dimexpr-\algorithmicindent-2pt\relax}\rule{88mm}{0.4pt}%
\Statex\hspace*{-\algorithmicindent}\textbf{#1}%
\vspace*{-.7\baselineskip}\Statex\hspace*{\dimexpr-\algorithmicindent-2pt\relax}\rule{88mm}{0.4pt}%
}

\newtheorem{property}{Property}

\newtheorem{lemma}{Lemma}

\long\def\longdelete#1{}

\begin{document}

\title{Link Delay Estimation Using Sparse Recovery for Dynamic Network Tomography}

\author{
    Hao-Ting Wei
\quad
    Sung-Hsien Hsieh
\quad
    Wen-Liang Hwang 
\\ \quad
    Chung-Shou~Liao 
\quad
    Chun-Shien Lu   

\thanks{
\noindent
H.-T. Wei and C.-S. Liao are with the Department
    of Industrial Engineering and Engineering Management, National Tsing Hua
    University,
    Hsinchu 30013, Taiwan.
S.-H, Hsieh, W.-L. Hwang, and C.-S. Lu are with the Institute of Information
Science,
Academia Sinica, Taipei 115, Taiwan.
}

\thanks{
\noindent
This work
was partially supported by
MOST 105-2221-E-007-085-MY3, 105-2628-E-007-010-MY3,
104-2221-E-001-019-MY3,
and 104-2221-E-001-030-MY3.
} }



\maketitle

\begin{abstract}
When the scale of communication networks has been growing rapidly in the past
decades,
it becomes a critical challenge to extract fast and accurate estimation of key
state parameters
of network links, {\em e.g.}, transmission delays and dropped packet rates,
because such monitoring operations are usually time-consuming.
Based on the \emph{sparse recovery} technique reported in
[Wang et al. (2015) {\em IEEE Trans. Information Theory}, 61(2):
1028--1044],
which can infer link delays from a limited number of measurements
using \emph{compressed sensing},
we particularly extend to
networks with
\emph{dynamic changes}
including
\emph{link insertion} and \emph{deletion}.
Moreover, we propose a more efficient algorithm with a better theoretical upper
bound.
The experimental result
also demonstrates that our algorithm
outperforms the previous work in running time
while maintaining
similar recovery performance,
which shows its capability to cope with
large-scale networks.

\end{abstract}

\begin{IEEEkeywords}
Compressed sensing, link delay estimation, network tomography, sparse recovery
\end{IEEEkeywords}


\vspace{-10pt}
\section{Introduction}



With the increasing trend of online services, including video conferencing and
live streaming, over the internet,
the need of high QoS (Quality-of-Service) guarantees has become urgent in recent
years.
The service providers are thus required to keep maintaining network utilization
and performance.
More precisely,
a network management system has to efficiently and repeatedly detect network
congestions and
estimate the status of link delays, if any exists.
The related works which discussed this problem, {\em i.e.},
\emph{network tomography} introduced by Vardi~\cite{V},
involved measuring origin-to-destination path delays
from aggregated measurements of the individual links along such paths in a given
network.
That is, these past studies on (\emph{active}) network tomography
measured end-to-end characteristics by sending probe packets from sender vertices
to
receiver vertices, {\em i.e.}, (actively) probing the
network~\cite{FR1,FR2,LMN,YCDV}.

Obviously, the computational load of visiting all link delays by measuring all
pairs of origin-to-destination paths is heavy, especially for large-scale
communication networks.
A common way to resolving this problem is a divide-and-conquer approach
which can reduce the size of a given network to smaller ones.
Hence, an initial estimation of state parameters
can be derived based on the principle of local information~\cite{YCDV}.
On the other hand, for its stochastic counterpart, where the link attributes
follow a given probabilistic distribution,
the EM (Expectation Maximization) algorithm was used to determine the unknown
state parameters.
Readers may refer to the survey~\cite{CCLNY} for related studies.
Recently, Firooz and Roy~\cite{FR1},
Mahyar {\em et al.}~\cite{MRH,MRMHGN},
Ghalebi {\em et al.}~\cite{GMGR},
Wang {\em et al.}~\cite{WXMT},
and Xu {\em et al.}~\cite{XH,XMA} proposed efficient approaches
to reducing the number of measurements by using the concept of \emph{compressed
sensing}.
Compressed sensing (CS)~\cite{CAN,DON} has been widely studied in the last decade,
which can be deemed as a combination of two phases:  \emph{sensing} and
\emph{signal recovery}.
The first phase is usually applied to data acquisition and compression (or
dimensionality reduction) simultaneously, while
the second phase aims to recover \emph{sparse signals}.
Here
we say a signal is $k$\emph{-sparse} if the number of nonzero entries of the
signal is at most $k$.
Based on the fact that there are usually a small number of links with large
delays,
the past studies that applied compressed sensing to network
tomography in that the parameters of link delays are represented by a \emph{sparse
vector}, and a measurement matrix by taking advantage of the network topology is
designed.

Compared with \cite{GMGR,MRH,MRMHGN,XMA}, our study and~\cite{FR1,WXMT} used a \emph{hub} set to measure the other links,
so that we can select the
set of remaining links randomly and this kind of measurement matrix has been proved to satisfy the RIP condition~\cite{BDD}.
Specifically, random matrices are known to satisfy $\delta_{ck}<\theta$ with high probability provided one chooses $m = O(ck/\theta^2\log(N/K) )$ measurements.
\cite{EJC} shows that $\ell_1$-minimization recovers all $k$-sparse vectors provided the sensing matrix satisfies $\theta_{2k}<0.414$.
The above theoretical results indicate that a sparse signal can be recovered from incomplete measurements under certain conditions.

More precisely, Wang et al.~\cite{WXMT} considered a model for
recovering sparse signals when satisfying some graph constraints.
This model finds applications to estimation of network parameters
because the parameters such as link delays, as mentioned, are usually sparse.
They adopted a \emph{line graph model} to represent the topological constraints to
ensure the network's connectivity,
where the constraints
allow that a subset of vertices can be aggregately measured
if and only if they induce a connected subgraph.
Moreover,
their two assumptions on the graph constraints are as follows:
\begin{enumerate}
\item[H1.] A vertex subset $V'$ of a graph $G$ can be measured together in
one measurement if and only if the subgraph induced by $V'$ is connected.
\item[H2.] The measurement is an additive sum of values at the corresponding
    vertices.
\end{enumerate}

\newpage

Here we follow the two assumptions
and our goal is to reduce the computation cost of transforming the network tomography problem
to a sparse recovery problem.
In particular, we extend it to
networks with
\emph{dynamic changes}, where
link insertion and deletion are allowed. That is,
we consider the problem model with dynamic graph constraints that
can be measured without recomputing all the state parameters under dynamic link
operations.
\vspace{-10pt}

\subsection{Main Contributions}

The key results obtained in this study are summarized as follows:

\begin{enumerate}
\item We consider
link delay estimation in a dynamic network tomography model,
where dynamic link operations
including
insertion and deletion are allowed.

\item We propose a faster algorithm and prove that it has a better theoretical
    upper bound
on running time, compared with Wang et al.'s work~\cite{WXMT}.

\item
The experimental result
confirms that our algorithm
also
outperforms~\cite{WXMT} in
running time, while maintaining similar
recovery performance.
Therefore,
it can be particularly applied to a large-scale network
environment.
\end{enumerate}



\longdelete
{
This paper is organized as follows:
In Section~2, we introduce problem models and some preliminaries.
In Section~3, we describe our algorithm for solving
the problem and show the theoretical analysis.
In Section~4, we introduce
the strategy
to cope with dynamic link
operations.
In Section~5, we show
numerical studies with simulation.
}
\vspace{-10pt}


\section{Problem Model}

We consider a communication network $G = (V,E)$, where $V$ denotes the set of
vertices with cardinality
$|V| = n$ and $E$ is the set of links with cardinality $|E| = m$.
Let $V = \{ v_1, v_2, \ldots, v_n \}$,
let $d_i$ be the degree of $v_i$,
and denote the average vertex degree as
$d= \frac{1}{n} \sum^n_{i=1} d_i$.
Wang et al.'s algorithm~\cite{WXMT}
adopted
a \emph{line graph model} $L_G=(V_L,E_L)$,
where every vertex in $L_G$ corresponds to a link in $G$, {\em i.e.}, $V_L=E$,
and two vertices are adjacent in $L_G$ if and only if
their corresponding links in $G$ are incident, {\em i.e.}, sharing a common end
vertex.
Similarly, let
$d'$ denote the average vertex degree in the line graph
$L_G$.

In \cite{WXMT},
every link in $E = \{ e_1,  \ldots, e_m \}$ is associated with a nonnegative delay
$x_i$,
$1 \le i \le m$, where $x=(x_1,\ldots,x_m)$ is the unknown signal to be recovered.
The authors assumed that
$x$ is a $k$-sparse vector; that is, the number of nonzero entries of $x$ is
at most $k$.
Based on that,
they can take $N$ measurements, $N \ll m$ for sparse signal recovery, where we let $y$ denote the measurement
vector of length $N$.
In addition, let $A$ be
the $N\times m$ binary measurement matrix,
where each row of $A$ represents a path
along which
entry $1$ (or $0$) denotes a
visited (or unvisited) link ({\em i.e.}, $A_{i,j} = 1$ if link $j$ contributes to the
$i$th measurement, and $A_{i,j} = 0$ otherwise).
Hence, one can have the compact form $y = Ax$.

Conventionally,
compressed sensing (CS) requires that each entry of $A$ be drawn from i.i.d random
variable (randomness constraint) for reconstructing $x$ given $y$ and $A$.
However, since G is not a complete graph, it leads to the difficulty in the design
of $A$.
To solve this problem, Wang et al.~\cite{WXMT} used the concept of a
connected dominating set
(CDS)\footnote{A set of vertices $C$ is a connected
dominating set of $L_G$, if every vertex in $L_G$ either belongs to $C$ or is
adjacent to a vertex in $C$, and $C$ is
connected.} on the line graph $L_G$ of a given graph $G$.
Let
$C$ denote a CDS on $L_G$.
Due to the property of a CDS,
every vertex
$v\in V_L\setminus C$
can be connected to
$C$.
That is,
we can keep connectivity by using the vertices in $C$
(as hub vertices) to
connect other vertices.
Moreover, we
can use the hub vertices to recover
others.
The above procedure is performed repeatedly until all the delays are
recovered.

For the implementation, a measurement is to
sum up all delays along a close path that
includes the hub
vertices and the delays
that
we want to measure.
Notice that
we have to deduct the delay of
the hub vertices
to
exactly
get the delay we want to measure.
More precisely,
assume the total delay is $y_t$ and the hub delay is $y_h$,
and let
what we want
to measure
be $y_t-y_h = (A_t -A_h)x$.
Therefore, $A_t-A_h$ is the
measurement array that
satisfies the randomness constraint~\cite{WXMT}.
The key idea of our algorithm
uses a similar manner,
i.e. the concept of CDS,
to ensure
the randomness constraint.
The major difference between our algorithm and Wang et al.'s approach is the
selection of \emph{hub} vertices; that is, finding a CDS in the initial round.


We further extend the model
to a \emph{dynamic} scenario
in which
link insertion and deletion are allowed.
Therefore, each input can be considered a sequence of instances, $I_1, I_2,
\ldots, I_T$, where
each instance $I_t$ is updated from $I_{t-1}$, $1 < t \le T$,
by inserting or deleting links in $G$.
Precisely,
every link $e_i$ is associated with a
$T$-dimensional vector, $x_i$, where each entry $x^t_i$
is derived from $x^{t-1}_i$
by inserting or deleting edge $e_i$ at time $t$.
That is,
$x^t = (x^t_1,x^t_2\ldots,x^t_m)$ denotes the unknown signal to recover
at time $t$, $1 \le t \le T$.
The objective is to recover and estimate the link delays at each time slot without repeating the whole procedure of selecting the set of hub vertices.

\vspace{-10pt}


\alglanguage{pseudocode}
\algrenewcommand\algorithmicrequire{\textbf{Input}}
\algrenewcommand\algorithmicensure{\textbf{Output}}

\section{Proposed Method}\label{Sec: Proposed Method}
Here we use the following topological property of
the original graph
to achieve a better selection of hub vertices.

\begin{property}
In a graph, a maximal independent set $S$ is also a dominating set; that is,
every vertex is either in $S$ or has at least one neighbor in $S$.
\end{property}

\begin{lemma}\label{lm1}
A matching in a graph $G$ corresponds to an independent set in the line graph
$L_G$ of $G$.
\end{lemma}
{\it Proof}:
Please see the supplementary material for proof.
\hfill $\Box$

\medskip

\longdelete{
{\it Proof}:
A matching $M$ is a set of edges that have no common end vertices.
When we consider the line graph $L_G$,
each vertex in $L_G$ corresponds to an edge in $G$. Because every two edges in $M$
are not incident in $G$,
in the line graph $L_G$ any two vertices in $M$ are not adjacent to each other.
$M$ is thus an independent set in $L_G$.
}
\begin{algorithm}
\begin{algorithmic}[1]
\setcounter{algorithm}{0}
\caption{\hspace{-1mm}: Construction of measurements for a graph
$G$}\label{our_algo}
\Require: $G=(V,E)$, $L_G=(V_L,E_L)$
\Ensure: All the measurements
\State Find a maximum matching $M$ in $G$;


\State Let $C^{\ast}=\textbf{Connecting}(G,M)$ and measure the summation delay of $C^{\ast}$; where $C^{\ast}$ corresponds to a CDS $C$ on $L_G$;

\State Design $f(|T|,k)+1$ measurements and the corresponding matrix to recover $x_T$,
where $T = V_L\setminus C$;
\State Measure the vertex delays
$x_C$ directly;



\end{algorithmic}

\hrulefill

{\bf Subroutine 1}: $Connecting(G,M)$
\begin{algorithmic}[1]
\Require: $G=(V,E)$, a maximum matching $M$
\Ensure: A connected edge set $C^{\ast}$

\State Let $C^{\ast}=M$, and randomly select a vertex $u$ ;


\While {$C^{\ast}$ is unconnected}

\State Find an edge $(u,v) \notin C^{\ast}$ and $(u,k),(v,r) \in M$;

\State $C^{\ast}: = C^{\ast} + (u,v)$;



\EndWhile
\end{algorithmic}
\end{algorithm}

Our algorithm is designed as follows and described in Algorithm 1.
We first find a
maximum matching in a given graph $G$,
and connect all the matched vertices to construct a connected edge set $C^{\ast}$
(see Subroutine 1).
Note that
the edge set $C^{\ast}$ corresponds to
a CDS $C$ in the line graph.

Based on~\cite{WXMT},
by letting every vertex in the CDS $C$ be a hub,
we can design an
appropriate number of measurements
to recover $x_{T}$ and derive $\hat{x_{T}}$ in the line graph $L_G=(V_L,E_L)$,
where
$T = V_L\setminus C$ and
$x_T$
is the subvector of
$x$ with
indices in $T$.
Here we let the appropriate number of measurements
be $f(|T|,k) +1$, where $f(|T|,k)$ represents
the number of measurements constructed to identify $k$-sparse vectors associated with a complete graph of $n$ vertices~\cite{WXMT}.
Then, we directly measure
each of the remaining delays in $x_C$ and obtain $\hat{x_{C}}$ , i.e.
$x_v$
with $v\in C$ in $L_G$.

We will discuss the strategy of dealing with
\emph{dynamic}
link operations
later in Section \ref{Sec: Dynamic Strategy}.
\vspace{-10pt}

\subsection{Analysis and Time Complexity}
Both of our algorithm and Wang et al.'s approach~\cite{WXMT}
recover link delays by using the concept of CDS
(i.e., hub vertices).
That is, in order to compare
our algorithm
with
that reported in \cite{WXMT}
from a theoretical perspective,
we
mainly consider the size of the
CDS, i.e.
the cardinality of the set of hub vertices
$C$,
as well as its construction time in the initial round.
Recall that Wang et al.
first calculated the radius
in
the line graph and
then picked
the central vertex to be the root vertex. Secondly, they used the breadth-first
search (BFS) and
treated the
non-leaf
vertices
as
the hub vertices.
The whole process of
their procedure to
generate a set of hub vertices costs $O({|E|^2}\log|E|+|E||E_L|)$
time in total~\cite{WXMT}.

\longdelete{ 
Suppose
every vertex in our network $G$ has probability $p$ to connect other vertices,
in the phase 1 of our method.
}

Given a network $G=(V,E)$, where $|V| = n$ and $|E|=~m$,
the cardinality of a maximum matching is
at most
$n/2$ so
the number of edges connecting
the maximum matching to form $C^{\ast}$
is at most $(n/2)-1$,
which implies that
the cardinality of $C^{\ast}$
is bounded by
$n$
in the initial round.
By contrast, the next lemma shows that the
number of
the hub vertices, generated by the BFS
method in~\cite{WXMT},
is $O(n \log n)$ in the worst case.

\begin{lemma}\label{lm2}
The cardinality of the hub selected by Wang et al.~\cite{WXMT} is
at least $n\log d$,
where $d$ is the average degree in a given network $G=(V,E)$.
\end{lemma}
{\it Proof}:
Please see the supplementary material for proof.
\hfill $\Box$

\medskip

\medskip

Consider the time complexity of Algorithm~\ref{our_algo},
the time cost mainly
comes from the computation of the maximum matching of a graph.
Since
Subroutine~1 can be solved at most linear in the number of vertices,
Micali and Vazirani~\cite{MV}
showed that the maximum matching problem can be solved in $O(|E| |V|^{0.5})$ time,
which
is much faster than \cite{WXMT} that takes at least
quadratic time proportional to the number of edges in a given network~$G$.
Hence, Algorithm~1 is more capable of
coping with large-scale networks.


\begin{algorithm}
\begin{algorithmic}[1]
\caption{: Dynamically maintain a connected edge set $C^{\ast}$}\label{dynamic_algo}
\Require: $G=(V,E)$, a maximum matching $M$, and a connected edge set $C^{\ast}$
\Ensure: An updated connected edge set $C^{\ast}$

\Algcase{Link-insertion of $(i,j)$}
\State Search for an augmenting path w.r.t $M$ in the graph
\Statex $G := G + (i,j)$;
\If {there exists an augmenting path}
\State Update the new maximum matching to be $M^*$;
\State
$C^{\ast}=\textbf{Connecting}(G,M^*)$;

\Else
\State $C^{\ast}$ is unchanged;
\EndIf

\Algcase{Link-deletion of $(i,j)$}

\State Search for an augmenting path w.r.t $M$ in the graph
\Statex 
$G := G
- (i,j)$;

\If {$(i,j)\in M$ and there is an augmenting path for $M - (i,j)$}
\State Update the new maximum matching to be $M^*$;
\State $C^{\ast}=\textbf{Connecting}(G,M^*)$;

\ElsIf {$(i,j)\notin M$ but $(i,j)\in C^{\ast}$ }
\State $C^{\ast}=\textbf{Connecting}(G,M)$;

\Else
\State $C^{\ast}$ is unchanged;
\EndIf
\end{algorithmic}
\end{algorithm}
\vspace{-10pt}


\begin{table*}[ht]
\renewcommand{\arraystretch}{1}
\begin{minipage}[c]{0.5\linewidth}
\caption{
Average time of selecting the set of hub vertices using
our and
Wang et al.'s algorithms in a scale-free network with different vertices ($n=500,1000$) and edge
degrees ($d=10,20$)
}\label{table1}
\centering
\begin{tabular}{|c|c|c|c|c|}
\hline
& Net 1 & Net 2 & Net 3 & Net 4 \\
  & $n=500$& $n=500$& $n=1000$& $n=1000$ \\
  &$d=10$ &$d=20$ & $d=10$&$d=20$\\
\hline
Wang et al's~\cite{WXMT} & $>10000 s$& $-$& $-$& $-$\\
\hline
Our algorithm &$4.7s$ &$5.79s$ &$28.15s$ &$43.22s$ \\
\hline
\end{tabular}
\end{minipage}
\vspace{-5pt}
\begin{minipage}[c]{0.5\linewidth}
\caption{Average time of selecting the set of hub vertices using
our algorithms in different scale-free networks with vertices ($n=500,1000$) and edge
degrees ($d=10,20$)
}\label{table2}
\centering
\begin{tabular}{|c|c|c|c|c|}
\hline
& Net 5 & Net 6 & Net 7 & Net 8 \\
  & $n=500$& $n=500$& $n=1000$& $n=1000$ \\
  & $d=10$ &$d=20$& $d=10$& $d=20$\\
\hline
Re-running & $4.70 s$& $5.79 s$& $28.15 s$& $43.22 s$\\
\hline
Dynamic strategy &$0.25  s$ &$0.27 s$ &$0.66 s$ &$ 0.69 s$ \\
\hline
\end{tabular}
\end{minipage}
\vspace{-15pt}
\end{table*}

\section{Dynamic Strategy}\label{Sec: Dynamic Strategy}

In this section, we
introduce our strategy to deal with dynamic link operations,
including
link insertion and deletion.
First, assume the given network remains connected when link deletion is allowed.
The dynamic procedure is described by Algorithm~2.
For link insertion of edge $(i,j)$,
we first find an augmenting path
by Micali and Vazirani~\cite{MV},
if any exists,
with respect to the original maximum matching $M$
in the graph
$G \cup \{(i,j)\}$.
Then, we derive the new maximum matching $M^*$ and
connect all the vertices in $M^*$ to form a new
set $C^{\ast}$
using Subroutine 1.
On the other hand, for link deletion, we
consider the following cases.
If $(i,j) \in M$,
we still use Subroutine 1 to
obtain a new $C^{\ast}$.
Otherwise,
if $(i,j) \notin M$,
that is, $M$ is still the maximum matching, and
if $(i,j) \in C^{\ast}$,
we can use Subroutine 1 to derive a new $C^{\ast}$.
Otherwise,
if $(i,j) \notin C^{\ast}$,
the original $C^{\ast}$ is optimal and $C^{\ast}$ does
not need to be
updated.
As shown in Algorithm~\ref{dynamic_algo},
the time cost of tackling dynamic link operations, {\em i.e.},
finding an augmenting path,
is at most linear in the number of edges.
In addition,
Subroutine~1
takes
at most linear time in the number of vertices.
Thus, the time complexity $O(n^2)$ of our dynamic strategy is faster than the one $O(n^{2.5})$ of re-running our proposed algorithm, where ``re-running'' means that there are dynamic link operations but we execute Algorithm~\ref{our_algo} without using the dynamic strategy.
\vspace{-10pt}



\section{Experimental Result}
We conducted numerical experiments
on the machine with Intel i5-6500(3.20GHz)
of CPU with
16GB DDR3 of RAM, running Windows 10 x64.
All programs were compiled by MATLAB R2015a and running time
was measured using one single thread.

There are two major issues we consider
for link delay estimation via sparse recovery:
execution time  and recovery performance.
\longdelete{
We evaluated the time cost of our algorithm in the following two scenarios.
We first considered a random network with $50$ vertices under different edge
probabilities with $p$ = $0.3$, $0.5$, $0.7$ and $0.9$ (see Table~\ref{table1})

Next we performed the algorithms in a random network with
a fixed
edge probability $p = 0.6$
under different numbers of vertices, i.e., $|V|$ = $20$, $40$, $60$
and $80$ (see Table~\ref{table1}).
}
We evaluated the time cost of our algorithm from the following perspectives.
We first considered
the B-A model\footnote{It is a kind of of scale-free networks and has also been proved that its degree distribution follows a power law, which corresponds to the environment of real-world networks.}~\cite{AB,WXMT},
where it is a scale-free network set to contain $500$ and $1000$ vertices under different average degrees $d$ = $10$ and $20$ (see Table~\ref{table1}), respectively.
we compared the time costs of re-running our algorithm and adopting the proposed dynamic strategy with edge deletion randomly
(see Table~\ref{table2}).


Obviously, the experimental result demonstrates that our algorithm outperforms
Wang et al.'s approach~\cite{WXMT} in terms of running time.
Precisely, our algorithm
runs
at least
thousand
times faster than Wang et al.'s
approach.
Moreover, we have
the following
observations from Table~\ref{table1} and Table~\ref{table2}:
1) As shown in Table~\ref{table1},
the execution time of Wang et al.'s algorithm is more than
ten thousand seconds.
\longdelete{
increases with the number of edges
due to its quadratic time complexity.
}
By contrast,
when the size of input networks grows,
there is only a
slight increase
in running time of
our
algorithm
because our algorithm takes $O(|E| |V|^{0.5})$ time only.
2) Table~\ref{table2} depicts the effectiveness of our algorithm to
deal with dynamic operations.
\longdelete
{
perform in large-scale
networks.
}
It is observed that our algorithm with dynamic strategy runs ten times faster
than its static version of re-running the algorithm without relyiong on the dynamic strategy.
Moreover,
when increasing
the size
of the scale-free network,
the time cost of the dynamic strategy increases slightly, which is significantly smaller than the increase cost of the static version.



As for the recovery performance, we
refer to the standard evaluation procedure for sparse recovery~\cite{CPL,JOJ}
and define
a procedure to be
\emph{success}
when all supports are found and $\frac{||\hat{x}-x_0||_2}{||x_0||_2} <0.02$.
Consider
the recovery performance
in a given scale-free network with
$|V|=500$ and
$d = 10$.
We generated a vector $x_0$ with a random support set,
where the sparsity rate, $r := k/m$, increases by $0.05$ from $0.05$ to $0.35$.
Next, the values of entries
in
the support set were drawn from a uniform distribution
with the range $[5(1-r),~5]$ and
others entries were drawn from the one
with range $[0,~0.001(1-r)]$.

\longdelete
{
We generated a
vector $x_0$
by randomly selecting supports, where
each support is multiplied by
a constant $ > 1$
and all the remaining ones are
multiplied by
a constant $\ll 1$,
in order to distinguish the delays}
We set the measurement ratio $N/m$ to be
$0.3$,~$0.4$ and $0.5$.
As shown in Fig.~1,
the recovery performance of our algorithm
outperforms Wang {\em et al.}'s method under low to moderate sparsity rates.
For example, when $N/m = 0.5$, our algorithm dominates their results until the sparsity is up to $25\%$.
In addition, we also
evaluated
the recovery performance of our algorithm with two different network densities.
As shown in Fig.~2, we verify our algorithm in two different scale-free networks with
$|V|=500$, and
$d = 10$ and $20$, respectively.
It can be
observed that,
under the same sparsity rate,
our algorithm exhibits better recovery performance in a more dense network.

More precisely, recall that $N = f(|T|,k) + |C|$, where $f(|T|,k)$ is the number of measurements to reconstruct $x_{T}$ and $C$ is the hub that is fixed and determined by the number of vertices of the input graph.
When the density of the network is increased, the ratio of $f(|T|,k)/N$ is increased correspondingly, leading to the advantage of more measurements for recovering $x_{T}$.
Therefore, our method has better recovery performance in a more dense network.

\begin{figure}[t]
\begin{minipage}[b]{0.7\linewidth}
\centering{\epsfig{figure=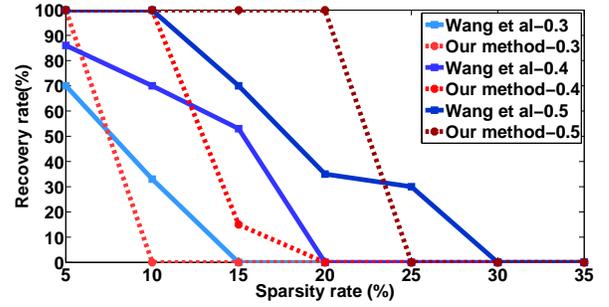,width=3.5in}}
\end{minipage}
\caption{The recovery performance of two methods with measurement ratios
$N/m$ = $0.3$,~$0.4$ and $0.5$}
\label{fig:Performance1}
\end{figure}


\begin{figure}[t]
\begin{minipage}[b]{0.7\linewidth}
\centering{\epsfig{figure=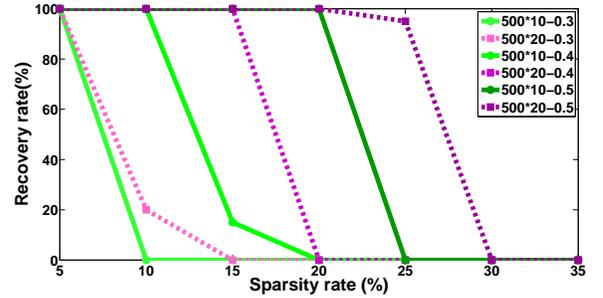,width=3.5in}}
\end{minipage}
\caption{Recovery performance of our method with respect to two different network densities under measurement ratios
$N/m$ = $0.3$,~$0.4$ and $0.5$}
\label{fig:Performance1}
\end{figure}
\vspace{-10pt}

\section{Conclusion}
This study
has extended
Wang {\em et al.}'s work~\cite{WXMT}
to link delay estimation with dynamic link operations.
Moreover, our algorithm ensures superior performance against the previous approach in
execution time while maintaining similar recovery performance.
Since our algorithm has significantly reduced the time expense, it is
well-expected
to be
implemented in large-scale networks.

There are
several
open
discussions
for this study.
Note that both of our algorithm and Wang {\em et al.}'s approach~\cite{WXMT}
attempted to recover link delays
by using
the concept of
connectivity.
\longdelete{
That is,
even the hub set in the initial round actually accounts for
a small proportion of the total amount of links,
which may cause the extra measurements to detect the delays.
It would be of great interest to design an
algorithm
to detect the whole link delays at one time.
}
It would be worthwhile to develop another
graph-theoretic approaches to effectively identify a hub.
Moreover, in this study, our dynamic strategy allows deleting one edge in a time.
Multiple edges deletion (or insertion) is an interesting issue for further study.

%

\clearpage
 \pagenumbering{arabic}
\section*{Supplementary material}

In this section, we present the proofs of the following lemmas in Sec. \ref{Sec: Proposed Method}.

{\it Lemma~\ref{lm1}:}
A matching in a graph $G$ corresponds to an independent set in the line graph
$L_G$ of $G$.

{\it Proof}:
A matching $M$ is a set of edges that have no common end vertices.
When we consider the line graph $L_G$,
each vertex in $L_G$ corresponds to an edge in $G$. Because every two edges in $M$
are not incident in $G$,
in the line graph $L_G$ any two vertices in $M$ are not adjacent to each other.
$M$ is thus an independent set in $L_G$.

\medskip

{\it Lemma~\ref{lm2}:}
The cardinality of the hub selected by Wang et al.~\cite{WXMT} is
at least $n\log d$,
where $d$ is the average degree in a given network $G=(V,E)$.

{\it Proof}:
Based on the analysis reported in~\cite{MAH},
the number of non-leaf vertices in a BFS tree is close to $\frac{n\ln d}{d}$.
Therefore, the cardinality of the hub approximates $\frac{m\ln d'}{d'}$ in
the corresponding line graph model $L_{G}(V_L,E_L)$ of $G$, where
$|V_L|= m$, $|E_L|= \frac{\sum_{i=1}^{n} {d_i}^{2}-2m}{2}$, and $d'$ is the average
degree of $L_{G}$.
Note that because
$|E_L| = \frac{d'\times m}{2} = \frac{\sum_{i=1}^{n} {d_i}^{2}-2m}{2}$, it implies
$d' = \frac{\sum_{i=1}^{n} {d_i}^{2}-2m}{m} = \frac{2\sum_{i=1}^{n}
{d_i}^{2}}{nd}-2$.\\
By the Cauchy-Schwarz inequality, we can derive: \\
\begin{align*}
&(\sum_{i=1}^{n}{d_i}^{2})(\sum_{i=1}^{n}{1}^{2})\geq (\sum_{i=1}^{n}{d_i})^{2}\\
& \Rightarrow\sum_{i=1}^{n}{d_i}^{2} \times n \geq {(nd)^2},
\end{align*}
which implies
$d' \geq \frac{2n{d^2}}{nd}-2 = 2d-2$.
Thus, the cardinality of the hub is close to
$\frac{m\ln d'}{d'} \leq~\frac{nd\ln (2d-2)}{4d-4}$,
which approximates $n\log d$.
Notice that the number of hub vertices selected by our approach is at most $n$.
In contrast, the cardinality of the hub derived by~\cite{WXMT},
$n\log d$, depends on the value of $d$, which ranges from $n$ to $n\log n$.

\end{document}